\documentclass[10pt,prd,aps]{revtex4}

\usepackage{graphics}
\usepackage[pdftex]{graphicx}

\newcommand{\ket}{\rangle}
\newcommand{\bra}{\langle}

\begin{document}

\title{The Discretised Adiabatic Theorem 
}
\vspace{-.99315cm}
\author{Bernhard  K. Meister}
\email{Bernhard.K.Meister@gmail.com}
\affiliation{ Department of Physics, Renmin University of China, Beijing, China 100872}

\date{\today }
\vspace{-.415cm}
\begin{abstract}

\noindent
A discretised version of the adiabatic theorem is described  with the help of a  rule relating a Hermitian operator to its expectation value and variance.
The simple initial operator $X$  with known ground state is transformed   in a series of $N$ small steps into a more complicated final operator $Z$ with unknown ground state. 
Each operator along the  discretised path in the space of Hermitian matrices  is used to measure the state, initially the  ground state of $X$. 
 Measurements similar to the Zeno effect or Renninger's negative measurements modify the state incrementally.  This process 
   eventually leads  to an eigenstate combination of  $Z$.   In the limit of vanishing step size the state  stays with overwhelming probability  in  the ground state of each of the $N$ observables. 


\end{abstract}
\maketitle
\vspace{-1.299015cm}
\subsection{Introduction}
\label{sec:1a}
\vspace{-.415cm}
\noindent
The quantum adiabatic theorem  was originally proposed in 1929 by  Born and Fock \cite{born} and exists in a variety of formulations.  The idea is to transform slowly, either continuously or, which is done in this paper, in a large number of discrete steps,  an  operator $X$ into $Z$ 
such that the ground state of  $X$  will stay with overwhelming probability in the ground state of the  operators along the path  and finally of  $Z$. 
This requires various assumptions about the spectrum of the operators  and the trajectory, and has been the subject of extensive research over the intervening decades.
For an introduction see a  text book on quantum mechanics like Messiah \cite{messiah}.

The paper proposes to apply a number of measurements   to a system starting with the ground state of $X$ to push it slowly towards an eigenstate of  the final operator.  The assumption of  non-degeneracy of the ground states of the observables along the whole path as well as an infinitesimal step size  will ensure  that the final state is the ground state of $Z$. In between measurements for sake of simplicity the state is assumed to be unchanged, i.e. the time  evolution of the system  is governed by a Hamiltonian given by the identity matrix.  A paper by Kitano\cite{kitano}, as well as others citing that work, also discusses the relationship between the (inverse) Zeno effect and  the adiabatic theorem, but uses a different method. 
Adding `inverse' to the term Zeno effect is apt, because instead of retarding the change of the state one facilities the change of the state through measurements. 
Renninger's negative measurements\cite{renn} modify the state incrementally in a way not dissimilar to the Zeno effect.
Only  the simplest case
 is considered, since the aim of the paper is to introduce a new method and not to explore applications. 
 
The structure of the paper is as follows. In the second section a statement about Hermitian operators is presented. In the subsequent section results for eigenvalues and eigenstates of a slowly varying observable are discussed. These results are in the final section applied to the adiabatic theorem. 
\vspace{-.615cm}
\subsection{A theorem for Hermitian operators}
\vspace{-.415cm}
\noindent
In this section a well-known result for Hermitian operators is presented. 
A Hermitian operator $A$ acting on any state $|\psi\ket$ can  be rewritten as a combination of the first and second moment,
\vspace{-.215cm}
\begin{eqnarray}
A|\psi\ket=\bra A\ket_{\psi} |\psi\ket + \sigma_{\psi}(A) |\psi^{\perp}_A\ket, \nonumber
\end{eqnarray}
for
\begin{eqnarray}
\bra A\ket_{\psi}:=\bra \psi |A|\psi\ket, \nonumber
\end{eqnarray}
and 
\begin{eqnarray}
 \sigma^2_{\psi}(A):=\bra \psi |A^2|\psi\ket- \bra A\ket_{\psi}^2, \nonumber
\end{eqnarray}
since
\begin{eqnarray}
 \bra \psi |A^{\dagger} A |\psi\ket =\big( \bra A\ket_{\psi}  \bra \psi| + \sigma_{\psi}(A) \bra \psi^{\perp}_A|\big)\big(\bra A\ket_{\psi} |\psi\ket + \sigma_{\psi}(A) |\psi^{\perp}_A\ket\big) =\bra A\ket_{\psi}^2+ \sigma^2_{\psi}(A). \nonumber
\end{eqnarray}
In addition,
\begin{eqnarray}
2 \sigma^2_{\psi}(A)=\sum_{i,j=1}^M |\alpha_i|^2 |\alpha_j|^2 (a_i - a_j )^2, \nonumber
\end{eqnarray}
for 
\begin{eqnarray}
|\psi\ket:=\sum_{i=1}^{M} \alpha_i |\psi_i\ket, \nonumber
\end{eqnarray}
where 
 the $|\psi_i\ket$ are an orthonormal basis of a $M$-dimensional Hermitian operator $A$ with eigenvalues $a_i$ and,
 $\sum_{i=1}^M |\alpha_i|^2 =1$.
In the next section the result above is applied to a progressively adjusted operator.
\vspace{-.615cm}
\subsection{Eigenstates and Eigenvalues of a slowly adjusting Operator}
\vspace{-.515cm}
\noindent
The eigenvalue and eigenvector of the operator $A$ modified by the addition of a Hermitian operator $\epsilon B$ are studied. 
For $|a\ket$ being an eigenstate of $A$ one gets
\begin{eqnarray}
(A+ \epsilon B)(|a\ket+\epsilon |c\ket) = \bra A\ket_a |a\ket + \epsilon \Big(\bra B\ket_a |a\ket + \sigma_a(B) |a_B^{\perp}\ket +\bra A\ket_c |c\ket + \sigma_c(A) |c_A^{\perp}\ket \Big)  + O(\epsilon^2)|d\ket , \nonumber
\end{eqnarray}
where $\epsilon$ is a small real number.
Normalised states with the following property
\vspace{-.015cm}
\begin{eqnarray}
\bra a| a\ket  = \bra a^{\perp}_B| a^{\perp}_{B}\ket 
=\bra c| c\ket=\bra c^{\perp}_A| c^{\perp}_{A}\ket=1\nonumber
\end{eqnarray}
 are used with  $|c\ket = \alpha |a\ket + \beta |a^{\perp}_C\ket$ and $| \alpha|^2+|\beta|^2=1$. In addition, $|d\ket$ is an element of the same Hilbert space.
If $|a\ket+\epsilon k |c\ket +O(\epsilon^2)|f\ket$ is an eigenstate of $A+ \epsilon B$, then
\vspace{-.015cm}
\begin{eqnarray}
(A+ \epsilon B)(|a\ket+\epsilon k |c\ket +O(\epsilon^2)|f\ket) = (\lambda+ \epsilon \gamma +O(\epsilon^2))(|a\ket+\epsilon k|c\ket +O(\epsilon^2)|f\ket) 
\nonumber
\end{eqnarray}
with $|f\ket$  an element of the Hilbert space and $k$ a complex number.
As a consequence, one can coral all the terms of order $\epsilon^0$ to get   $ \lambda=\bra A\ket_a$, and
similarly all the terms of order $\epsilon^1$ to get $     \bra A\ket_a  k|c\ket+ \gamma  |a\ket  =\bra B\ket_a |a\ket + \sigma_a(B) |a_B^{\perp}\ket +k \bra A\ket_c |c\ket + k\sigma_c(A) |c_A^{\perp}\ket $.
This can be rewritten as 
\begin{eqnarray}
k\big(\bra A\ket_a - \bra A\ket_c\big) |c\ket =\big( \bra B\ket_a -\gamma \big) |a\ket  + \sigma_a(B) |a_B^{\perp}\ket + k\sigma_c(A) |c_A^{\perp}\ket, \nonumber
\end{eqnarray}
\vspace{-.065cm}
and furthermore 
\vspace{-.065cm}
\begin{eqnarray}
&&|(\bra a|+\epsilon k^* \bra c|   + O(\epsilon^2) \bra f|)
(|a\ket +\epsilon k |c\ket)+O(\epsilon^2)|f\ket|^2=(1+\epsilon k \alpha)(1+\epsilon k^*\alpha^*)+O(\epsilon^2 )
=1 +  2 Re( \alpha k) \epsilon+O(\epsilon^2 ). 
\nonumber
\end{eqnarray}
\vspace{-.065cm}
 The transition probability between $|a\ket$ and  the normalised version of $|a\ket +\epsilon k|c\ket$ is
\vspace{-.065cm}
\begin{eqnarray}
\frac{|(\bra a|+\epsilon k^* \bra c|+O(\epsilon^2) \bra f|)
|a\ket |}{\sqrt{1+2Re( \alpha k) \epsilon+O(\epsilon^2)}}&=&\frac{1+Re( k \alpha)\epsilon+O(\epsilon^2) }{\sqrt{1+2Re( k\alpha )\epsilon+O(\epsilon^2)}}\nonumber\\
&=&1+O(\epsilon^2).\nonumber
\end{eqnarray}
Now as one applies this procedure $1/\epsilon$-times, or equivalently $N$-times with $1/\epsilon=N$, one gets as the product of overlaps
\vspace{-.065cm}
\begin{eqnarray}
 \lim_{\epsilon\to 0}\Big(1+ O(\epsilon^2)\Big)^{1/\epsilon}
 = \lim_{N\to \infty}
 \Big(1+O(N^{-2}) \Big)^{N}
 =1,\nonumber
\end{eqnarray}
since $\lim_{N\to \infty} N\log(1+const /N^2+ O(1/N^3)) =\lim_{N\to \infty} N (const /N^2) + O(1/N^2) = 0$.
In the next part the result is applied to the adiabatic theorem, since one can in a series of steps with the associated measurements move from one operator to any other other operator and push the initial ground state to the final ground state. 
\vspace{-.615cm}
\subsection{Discrete Adiabatic Theorem in the limiting case of $N\rightarrow \infty$}
\vspace{-.415cm}
\noindent
In this section the results of the previous two  are applied to a particular sequence of $N$ observables, starting at $X$ and ending up at $Z$.
To do this we consider a sequence of operators $A_j$ of the following form
\vspace{-.065cm}
\begin{eqnarray}
A_{j+1}=A_j+B \epsilon=(1-j \epsilon) X+j \epsilon Z \,\,\,\,     {\rm with}\,\, \epsilon=1/N \,\, {\rm and} \,\, \forall j\in\{0,1,..,N\}.\nonumber
\end{eqnarray}
with $A_0=X$, $B=Z-X$.
The result from the previous section can be applied directly. In the limit of $N$ going towards infinity the initial ground state - assuming no degeneracy in the spectrum of the $A_j$ along the way - is forced  into the final ground state at the end of the process.

Next, we calculate the ground state transition probability between subsequent measurement operators $A_j$ and $A_j+\epsilon B$.
The lowest level eigenstate of $A_j$ for a specific $j$ is assumed to be $|a\ket$ and the ground state of $A_j+\epsilon B$ is defined to be $|d\ket$.
Let us expand $|d\ket$ in terms of powers of $\epsilon$
\vspace{-.065cm}
\begin{eqnarray}
|d\ket = \sum_{i=0}^{\infty} \epsilon^i |d_i\ket, \nonumber
\end{eqnarray}
where $|d_i\ket$ are not necessarily normalised or orthogonalised  states. The derivation only slightly differers from the one given above.
Since in the limit $\epsilon \rightarrow 0$, $|d\ket$ will coincide with $|a\ket$, $|d_0\ket$ is equal to $|a\ket$.
The norm of $|d\ket$  up to $O(\epsilon^3)$ is 
\vspace{-.015cm}
\begin{eqnarray} 
\bra d| d\ket = 1 + \epsilon (\bra a | d_1  \ket  + \bra d_1 | a \ket ) + \epsilon^2 ( \bra d_1 |d_1 \ket  + \bra a | d_2  \ket  + \bra d_2 | a \ket ))  +O(\epsilon^3).\nonumber
\end{eqnarray}
The normalised version of $|d\ket$ is called $|\hat{d}\ket$. The higher order terms need not to be considered to evaluate the limit. As a reminder, this is different, if one considers a more realistic case with a potential degeneracy of the ground state or infinitesimal separation between ground state and first excited state, where higher order terms of $\epsilon$ cannot be ignored without further justification. 
The inner product $\bra a| \hat{d}\ket$ is evaluated next:
\vspace{-.215cm}
\begin{eqnarray} 
\bra a| \hat{d}\ket = \frac{1 + \frac{1}{2}\epsilon (\bra a | d_1  \ket  + \bra d_1 | a \ket ) +  
\frac{1}{2}\epsilon^2 (  \bra a | d_2  \ket  + \bra d_2 | a \ket ))  +O(\epsilon^3)}{\sqrt{1 + 
\epsilon (\bra a | d_1  \ket  + \bra d_1 | a \ket ) + \epsilon^2 (  \bra d_1 |d_1 \ket  + 
\bra a | d_2  \ket  + \bra d_2 | a \ket ))  +O(\epsilon^3)}}= 1-  
\frac{1}{2}\epsilon^2 \Big( \bra d_1 |d_1 \ket -\frac{1}{2} (\bra a | d_1  \ket  +
 \bra d_1 | a \ket )^2 \Big)+O(\epsilon^3).\nonumber
\end{eqnarray}
If   measurements are carried out $N$-times  in sequence, with $\epsilon=1/N$,  such that each time the measurement operator is slightly adjusted, the result  in the limit is
\vspace{-.015cm}
 \begin{eqnarray} 
\lim_{N\rightarrow \infty}|\bra a| \hat{d}\ket|^N= \lim_{N\rightarrow \infty} \Big(1-  \frac{1}{2N^2} \Big( \bra d_1 |d_1 \ket -\frac{1}{2} \Big(\bra a | d_1  \ket  + \bra d_1 | a \ket \Big)^2 \Big)+O(N^{-3})\Big)^N
= 1.
\nonumber
\end{eqnarray}
\vspace{-.0815cm}
Each measurement projects the state onto the basis of the operator to be measured. In the limit of $N\rightarrow \infty$ with probability one  the initial ground state transforms into the final ground state, if no degeneracies of the ground state and associated crossings are allowed along the discretised path. 


This paper  introduced a new version of the adiabatic theorem in the discrete setting for $N\rightarrow \infty$ measurements.
Various open questions remain.
It is of interest to understand what happens in the case of a finite number of measurements each taking a non-vanishing time.  The aim is to limit the measurement number and duration, 
while still ending up at the ground state of the final operator with a reasonable high probability.   
How does the optimal individual measurement number and duration  depend on the energy level separation as well as the strength of the measurement field?  
This will be studied in a separate paper.
In contrast, the  aim of this note is   to discuss the general application of the result of section B to the simplest discrete version of the adiabatic theorem. 

Each of the series of sequential measurements can be compared to the application of a quantum gate as used in the circuit model of quantum computation. The measurement based approach is likely to be less efficient than the circuit model or adiabatic quantum computation, since the failure rate only scales linearly, as  $1/N$, with the number of steps.  

Geometric ideas might turn out to be useful to find the optimal path, since the trajectory from initial to well-defined final point in the space of Hermtian matrices can take many forms. An appropriate metric has to be chosen. 
A rather rough analogy is discussed next. Like a hike across a mountain side with definite start and end point, one should avoid treacherous parts or at least  after every  deadly plunge  modify ones route to try achieve the transition  in an acceptable time. 
Ideal would be a metric on the space of Hermitian observables with  a distance function that takes ground state separation to next eigenvalue into account and has a divergence, if  the ground state of one of the observables becomes degenerate\footnote{One possibility would be to set $d(X,Z)$ equal to  $|(|\alpha_1-\alpha_2|)^{-1}-(|\omega_1-\omega_2|)^{-1}|$, where the Eigenvalues of the 2-dimensional matrix $X$ are $\alpha_1$ and $\alpha_2$ and the eigenvalues of $Z$ are $\omega_1$ and $\omega_2$. The distance diverges, if one or both of the matrices have  degenerate eigenvalues.}.
  An optimal path  would avoid  difficult patches. 
If the state eigenvalue separation on the discrete steps of the trajectory does not meet the requirements, one can vary the result by adding a Brownian bridge. 
Randomness and switching between different initial states adds a liberating note of unpredictability into the process and prevents one from following the same path repeatedly.  If  too many attempts fail, then this suggests an unavoidable  degeneracy or at least a minute energy level separation between the ground state and first excited state somewhere along the path. This forms  in terms of the distance function  an unsurmountable  mountain ridge between the initial and  final operator.
A connection to adiabatic quantum computing with its heavy reliance on the adiabatic theorem can easily be established. Implications will be discussed  elsewhere.\\
%
%
%
\noindent
  Discussions  with D.C. Brody and L.P. Hughston are gratefully acknowledged.

\vspace{-.915cm}
\end{document}